\documentclass[aps,pre,reprint,groupedaddress,showpacs,preprintnumbers,amsmath,amssymb,superscriptaddress]{revtex4-1}

\usepackage{graphicx}
\usepackage{dcolumn}
\usepackage{bm}

\DeclareMathAlphabet{\mathsfsl}{OT1}{cmss}{m}{sl}

\begin{document}

\title{The effect of electrostatic shielding using invisibility cloak}

\author{Ruo-Yang Zhang}
\email{zhangruoyang@gmail.com}
\affiliation{
 Theoretical Physics Division, Chern Institute of Mathematics, Nankai University, Tianjin,  300071, P.R.China
}

\author{Qing Zhao}%
\email{qzhaoyuping@bit.edu.cn}
\affiliation{
Department of Physics, College of Science, Beijing Institute of Technology, Beijing, 100081, P.R. China
}

\author{Mo-Lin Ge}%
\email{geml@nankai.edu.cn}
\affiliation{
 Theoretical Physics Division, Chern Institute of Mathematics, Nankai University, Tianjin,  300071, P.R.China
}
\affiliation{
Department of Physics, College of Science, Beijing Institute of Technology, Beijing, 100081, P.R. China
}

\date{\today}

\begin{abstract}
The effect of electrostatic shielding for a spherical invisibility cloak with arbitrary charges inside is investigated. Our result reveals that the charge inside the cloak is a crucial factor to determine the detection. When charged bodies are placed inside the cloak with an arbitrary distribution, the electric fields outside are purely determined by the total charges just as the fields of a point charge at the center of the cloak. As the total charges reduce to zero, the bodies can not be detected. On the other hand, if the total charges are nonzero, the electrostatic potential inside an ideal cloak tends to infinity. For unideal cloaks, this embarrassment is overcome, while they still have good behaviors of shielding. In addition, the potential across the inner surface of an ideal cloak is discontinuous due to the infinite polarization of the dielectric, however it can be alternatively interpreted as the dual Meissner effect of a dual superconductive layer with a surface magnetic current.
\end{abstract}


\maketitle

\section{\label{sec:level1}Introduction}

Since the pioneering works of Pendry \textit{et al.} \cite{Pendry2006Sci} based on coordinate transformation and Leonhardt \cite{Leonhardt2006Sci} based on conformal mapping method, invisibility cloak has attracted much attention and widely research. Noteworthy, in 2003, Greenleaf \textit{et al.} have suggested to design anisotropic conductivities that cannot be detected by electrical impedance tomography through coordinate transformation of Poisson's equation which is similar to Pendry's method \cite{Greenleaf2003PM,Greenleaf2003MRL}. The effectiveness of transformation based cloak with passive objects inside has been verified through ray tracing approach \cite{Schurig2006OE, Niu2009OE}, full-wave simulations \cite{Cummer2006PRE}, and analysis based on scattering models \cite{Chen2007PRL, Ruan2007PRL, Zhang2007PRB, Luo2008PRB}. On the other hand, the case that active sources are located inside the hidden area has also been investigated both in physics \cite{Zhang2008PRL} and mathematical senses \cite{Greenleaf2007CMP}. These investigations revealed the ideal invisibility cloak at particular frequency can prevents the electromagnetic waves generated by the sources inside the cloak from propagating out, meanwhile an extra surface electric and magnetic voltage are induced by the infinite polarization on the inner surface. As a result, it seems that the invisibility cloak with passive or active objects inside can not be detected by means of electromagnetic waves, while another method was suggested to detect the cloak by shooting a fast-moving charged particle through it \cite{Zhang2009PRL}.

In previous researches, the hidden passive objects and active sources are all electric neutral. As far as we know, the situation that the hidden objects are charged has never been discussed. However, the charges inside the cloak would be an important factor influencing the detection of the hidden objects. In this paper, we will only discuss the situation that the distribution of charges does not change with time, so the shielding effect of the invisibility cloak reduces to an electrostatic problem. The electrostatic problem could be regarded as the limit of harmonically varying fields as $\omega\rightarrow 0$. However, some special property of harmonic fields could remain and the electric and magnetic fields would not decouple completely even at the limit $\omega\rightarrow 0$, such as $|E/B|\equiv c$ does not varying with $\omega$ for a plane wave propagating in vacuum. So we treat the problem through solving the Poisson's equation strictly rather than taking the limit of zero frequency. 

In this paper, we fist construct the electrostatic cloak according to the invariance of Poisson's equation under coordinate transformation. Actually, the method to design electrostatic cloak has been proposed by Greenleaf  \textit{et al.} \cite{Greenleaf2003PM,Greenleaf2003MRL}. However, Greenleaf's device
is made of conductivity , while the cloak suggested here is composed of dielectric. Then, we calculate the fields in the whole space where the charges with arbitrary distribution are located inside a spherical cloak which are constructed by arbitrary radial transformation $f(r)$. The result reveals the fields out of the cloak are only determined by the total charges inside the cloak and the shielding behavior of the cloak are very like the effect of charge confinement by a dual superconductor under the monopole-existing supposition, which model is often used in color confinement \cite{Ripka2005Springer}. Moreover, for ideal cloaks, the result suggests an infinite electric energy when the total charges are nonzero. To overcome the embarrassment, we investigate the unideal cloaks and find that they can also realize a good effect of shielding.

\section{Electrostatic cloak}

In vacuum, the Electrostatic fields obey Poisson's equation \cite{Greenleaf2003MRL, Greenleaf2003PM}:
\begin{equation}\label{Poisson equation}
  \nabla^2 \psi\ =\ \frac{1}{\sqrt{\gamma}}\partial_i(\sqrt{\gamma}\gamma^{ij}\partial_j \psi)\ =\ -\frac{\rho}{\varepsilon_0},
\end{equation}
where $\psi$ is the electrostatic potential, $\gamma^{ij}$ is the contravariant component of the spatial metric under an arbitrary curvilinear coordinate system (marked as S system) of the 3-D vacuous virtual space (VS), and $\gamma$ is the determinant of  $\gamma_{ij}$. In Eq.~(\ref{Poisson equation}), $\sqrt{\gamma}\gamma^{ij}$ can be alternatively interpreted as the permittivity $\varepsilon^{ij}$ of a real material expressed in another coordinate system of a real physical space (PS). Meanwhile, the physical quantities in VS also can be expressed in the same system as in PS, which is marked as S' system. Concerning an arbitrary radial transformation between  S' and S system: $r'=f(r)$, $\theta'=\theta$, $\phi'=\phi$, where S' is a spherical coordinate system, we obtain the relative permittivity of the spherical electrostatic cloak through the same process as in optical cloaking:
\begin{equation}\label{permittivity}
  \varepsilon_{\langle ij \rangle} =
  \mathrm{diag}\left(\frac{f^2(r)}{r^2f'(r)},\ \  f'(r), \ \ f'(r)
  \right).
\end{equation}
where we use $\varepsilon_{\langle ij \rangle}$ to denote the uncoordinate components of the permittivity in orthonormal bases.

We can see that the expression of permittivity of an electrostatic cloak is identical with it of an optical cloak \cite{Luo2008PRB}.
So the electrostatic cloak can be regarded as an optical cloak at zero frequency. We will verify that the spherical electrostatic cloak expressed in Eq.~(\ref{permittivity}) with ``invisibility conditions'' $f(a)=0$, $f(b)=b$ can not be detected in electrostatic sense, where $a,\ b$ are the inner and outer radiuses of the cloak respectively.

Consider an electrostatic cloak wrapping a homogenous medium $\varepsilon_1$ is located in vacuum endowed with an uniform electric field. A spherical coordinate system  is constructed whose origin is at the center of the cloak and $z$ axis is along the direction of the field. So the original field is $\psi_0=-E_0r\cos\theta+C$, where $E_0$ is the intensity of the uniform field. For charge-free anisotropic media, the potential satisfies $\nabla\cdot(\tensor{\varepsilon}\cdot\nabla\psi)=0$. Substituting Eq.~(\ref{permittivity}) into the formula, we have
\begin{equation}\label{Poisson equation in S'}
  \frac{\partial}{\partial f}\left(  f^2\frac{\partial\psi}{\partial f} \right)+\frac{1}{\sin\theta}\frac{\partial}{\partial \theta}\left( \sin\theta\frac{\partial\psi}{\partial \theta} \right)+\frac{1}{\sin^2\theta}\frac{\partial^2\psi}{\partial \phi^2} = 0,
\end{equation}
which is just the Poisson's equation in S' system of VS. Through separating variables, $\psi$ can be represented by series

\begin{subequations}\label{series}
\begin{flalign}\label{series1}
 \psi^{\mathrm{out}}\ =\ &\sum_{n=0}^{\infty}B^{\mathrm{out}}_n\frac{1}{r^{n+1}}P_n(\cos\theta),\\\label{series2}
 \psi^{\mathrm{cl}}\ =\ &\sum_{n=0}^{\infty}\left( A^{\mathrm{cl}}_n f(r)^n + B^{\mathrm{cl}}_n\frac{1}{f(r)^{n+1}}\right) P_n(\cos\theta),\\\label{series3}
 \psi^{\mathrm{int}}\ =\ &\sum_{n=0}^{\infty} A^{\mathrm{int}}_n r^n P_n(\cos\theta),
\end{flalign}
\end{subequations}
where $\psi^{\mathrm{out}}$, $\psi^{\mathrm{cl}}$, $\psi^{\mathrm{int}}$ represent the reflective field outside the cloak, the field in the cloak shell and the field inside the internal hidden area respectively, and $P_n(\cos\theta)$ denotes the $n$-order Legendre function.

According to the boundary  conditions, which is the continuity of $\psi$ and the normal component of $\vec{D}$ across the inner and outer surface of the cloak, we obtain
\begin{subequations}
\begin{gather}
  A^{\mathrm{cl}}_0 =  A^{\mathrm{int}}_0 = c,\\
  A^{\mathrm{cl}}_1 =  \alpha E_0,\quad B^{\mathrm{cl}}_1 =  \beta E_0,\\
  A^{\mathrm{int}}_1 =  \frac{1}{a}\big[\alpha f(a)+\frac{\beta}{f(a)^2}\big]E_0,\\
  B^{\mathrm{out}}_1 =  b^2\big[\alpha f(b)+\frac{\beta}{f(b)^2}+b\big]E_0,
\end{gather}
\end{subequations}
and other coefficients are all zero, where
\begin{widetext}
\begin{subequations}
\begin{flalign}
  &\alpha=\frac{3f(b)^2b^2(\frac{\varepsilon_1}{\varepsilon_0}a+2f(a))}{2f(a)^3(f(a)-\frac{\varepsilon_1}{\varepsilon_0}a)(f(b)-b)-f(b)^3(\frac{\varepsilon_1}{\varepsilon_0}a+2f(a))(2b+f(b))},\\
  &\beta=\frac{3f(a)^3f(b)^2b^2(f(a)-\frac{\varepsilon_1}{\varepsilon_0}a)}{2f(a)^3(f(a)-\frac{\varepsilon_1}{\varepsilon_0}a)(f(b)-b)-f(b)^3(\frac{\varepsilon_1}{\varepsilon_0}a+2f(a))(2b+f(b))}.
\end{flalign}
\end{subequations}
\end{widetext}
When $f(a)=0$, $f(b)=b$, they reduce to $\alpha=-1$, $\beta=0$. Therefore, $\psi^{\mathrm{out}}=0$, $\psi^{\mathrm{int}}=C$, and the field in the cloak layer is
\begin{equation}\label{potential in cloak}
\psi^{\mathrm{cl}}\ =\ -E_0f(r)\cos\theta+C.
\end{equation}
The result indicates the perfect cloak and its hidden medium can not be detected in that there is no reflective field. Furthermore, the hidden area can not perceive the fields outside the cloak either.

\begin{figure}[t]
\includegraphics[width=0.45\columnwidth,clip]{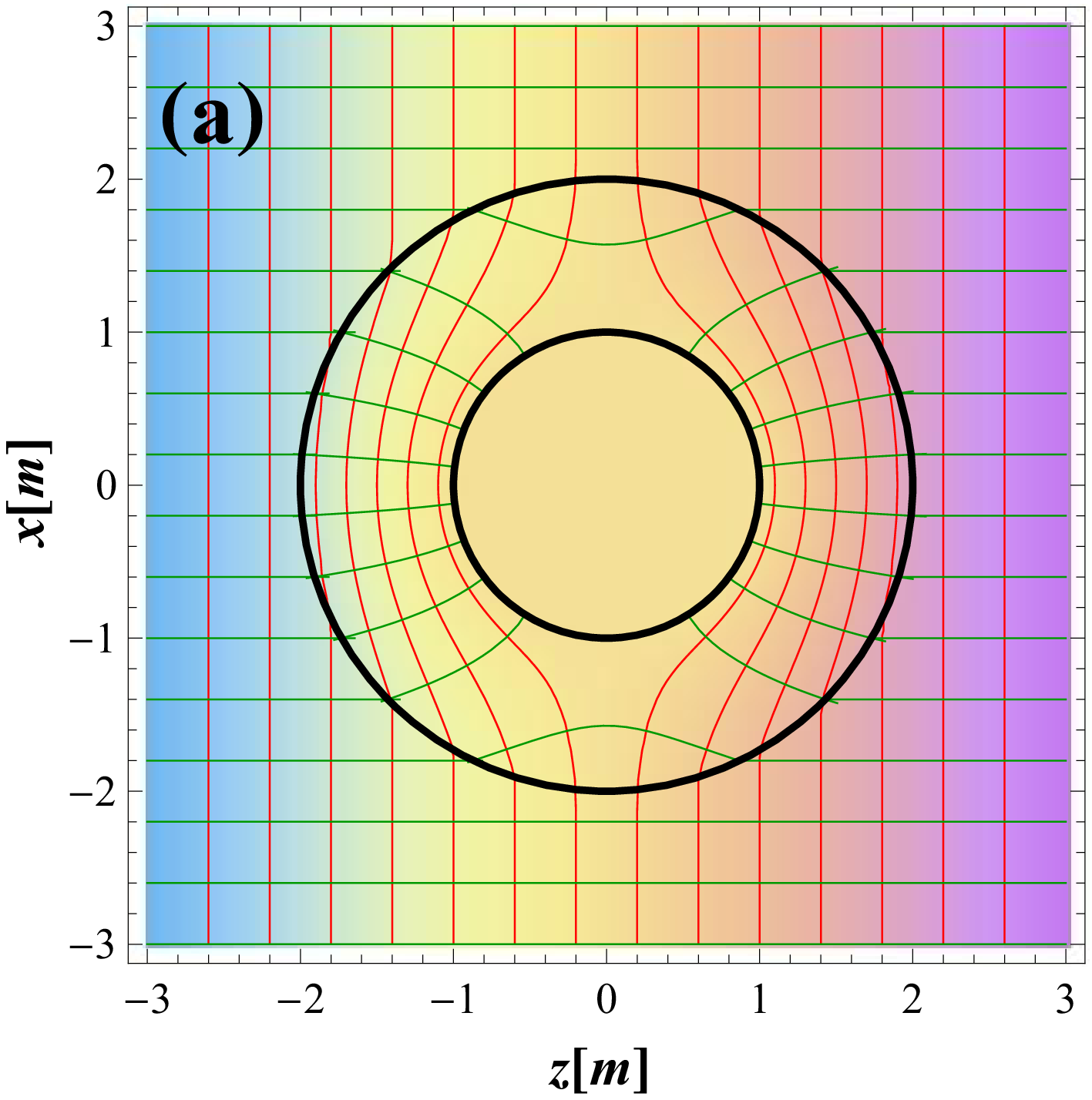}
\includegraphics[width=0.45\columnwidth,clip]{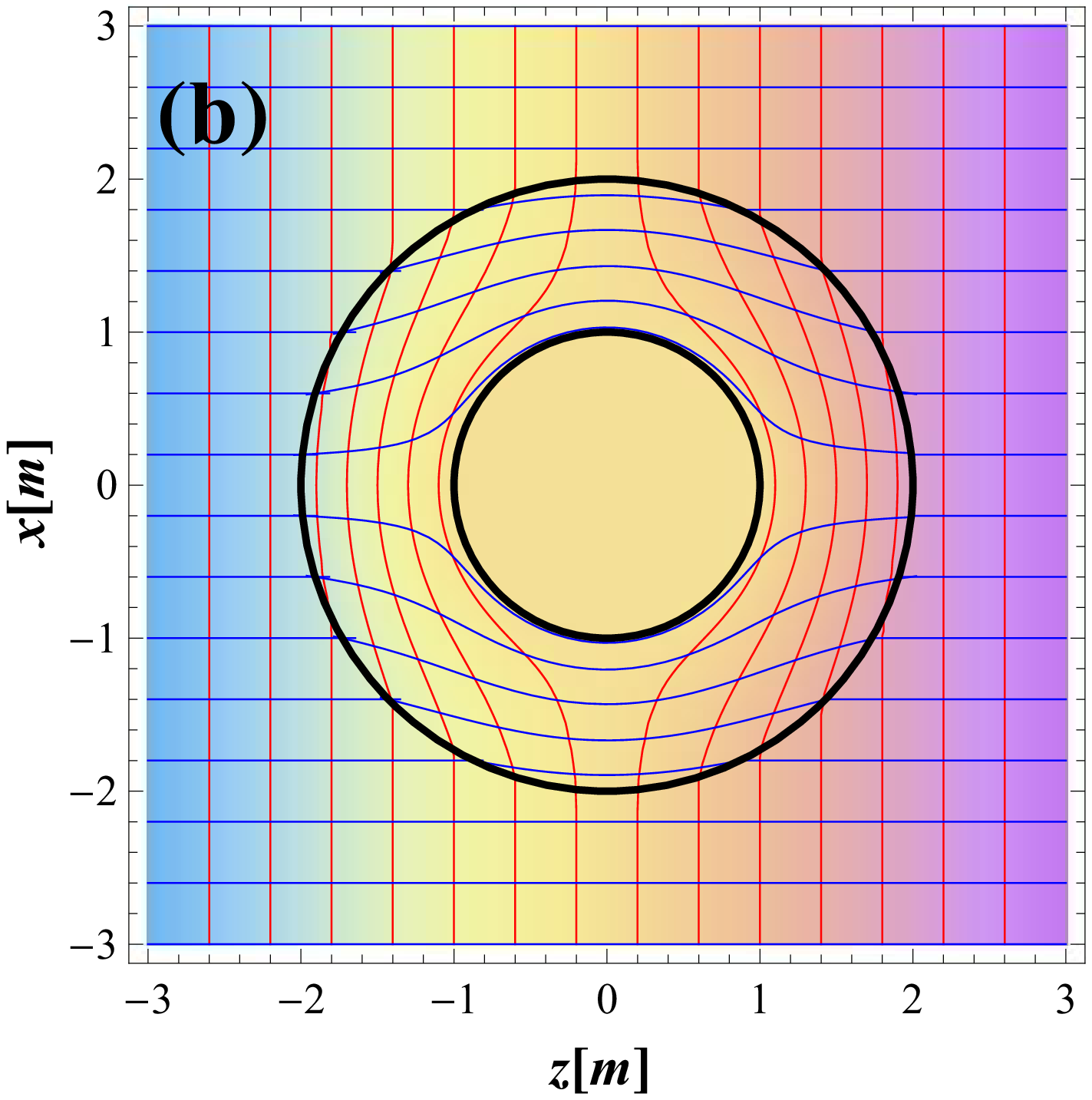}
\caption{\label{fig1}Electrostatic cloak in uniform electric field. The distribution of potential are illustrated in $y=0$ plane. Red lines denote equipotential surfaces, (a) green lines denote electric lines of force, and (b) blue lines denote lines of electric displacement. The inner and outer radiuses of cloak are $a=1\mathrm{m}$ and $b=2\mathrm{m}$ respectively.}
\end{figure}

In terms of Eq.~(\ref{potential in cloak}), we get the electric intensity and the electric displacement vector in the cloak layer:
\begin{subequations}
\begin{flalign}
  \vec{E}^{\mathrm{cl}}\ =\ &E_0\left[ f'(r)\cos\theta\hat{e}_r-\frac{f(r)}{r}\sin\theta\hat{e}_\theta \right],\\
  \vec{D}^{\mathrm{cl}}\ =\ &\frac{\varepsilon_0 E_0f(r)}{r}\left[ \frac{f(r)}{r}\cos\theta\hat{e}_r-f'(r)\sin\theta\hat{e}_\theta \right].
\end{flalign}
\end{subequations}
The electric field intensity can be regarded as the tangential vector of the electric lines of force, that is $d\vec{r}/d\lambda=\vec{E}^{\mathrm{cl}}$ where $\lambda$ is the parameter. Thus, we can derive the analytical expression of the electric lines of force
\begin{equation}\label{wave-normalinsphericalcloak}
   \exp\left[\int\frac{f(r)}{r^2f'(r)}dr\right]\sin\theta\ =\ \mbox{const},\quad \phi\ =\ \phi_0,
\end{equation}
which employ the same expression of wave-normal rays in optical cloak exposed in a plane wave. Similarly, the lines of electric displacement are expressed by
\begin{equation}\label{electric lines of force}
f(r)\sin\theta\ =\ \text{const},\quad \phi\ =\ \phi_0.
\end{equation}
It is just the expression of straight lines in S' system of VS, and it corresponds to the light-ray in optical cloak. The distribution of the fields are shown in Fig.~\ref{fig1}, where the constant of the potential are selected to be $C=0$.

\section{electrostatic shielding}

Now we discuss the problem of electrostatic shielding of the cloak with an arbitrary distribution of charges inside. The simplest situation is that a point charge lies on the $z$ axis with a displacement $a_0(<a)$ from the origin. The potential in each region holds the same form as in Eqs.~(\ref{series}) except that the total potential inside the hidden area should add the potential produced by the point charge
\begin{equation}
   \psi^{\mathrm{int}} = \frac{\tilde{q}}{\sqrt{r^2-2a_0r\cos\theta+a_0^2} }+\sum_{n=0}^{\infty} A^{\mathrm{int}}_n r^n P_n(\cos\theta),
\end{equation}
where $\tilde{q}=q/(4\pi\varepsilon_1)$ and $q$ is the quantity of the point charge. The zero-point of potential is selected at infinity. When $r>a_0$, the potential of point charge can be expanded as
\begin{equation}
\frac{\tilde{q}}{\sqrt{r^2-2a_0r\cos\theta+a_0^2} }= \tilde{q}\sum_{n=0}^{\infty} \frac{a_0^n}{r^{n+1}} P_n(\cos\theta).
\end{equation}
Substitution the potentials into the boundary conditions at $r=a$ and $b$ yields
\begin{subequations}
\begin{flalign}
  &\frac{B^{\mathrm{out}}_n}{b^{n+1}}-f(b)^nA^{\mathrm{cl}}_n-\frac{B^{\mathrm{cl}}_n}{f(b)^{n+1}}=0,\\
  &\frac{n+1}{b^{n+2}}B^{\mathrm{out}}_n+\frac{nf(b)^{n+1}}{b^2}A^{\mathrm{cl}}_n-\frac{n+1}{b^2f(b)^n}B^{\mathrm{cl}}_n=0,\\
  &f(a)^nA^{\mathrm{cl}}_n+\frac{B^{\mathrm{cl}}_n}{f(a)^{n+1}}-\frac{\tilde{q}a_0^n}{a^{n+1}}-a^nA^{\mathrm{int}}_n=0,\\
  \begin{split}
  &\frac{nf(a)^{n+1}}{a^2}A^{\mathrm{cl}}_n-\frac{n+1}{a^2f(a)^n}B^{\mathrm{cl}}_n+\frac{\varepsilon_1}{\varepsilon_0}\frac{(n+1)\tilde{q}a_0^n}{a^{n+2}}\\
  &-\frac{\varepsilon_1}{\varepsilon_0}na^{n-1}A^{\mathrm{int}}_n=0,
  \end{split}
\end{flalign}
\end{subequations}
By solving the equations, all coefficients are determined
\begin{subequations}
\begin{flalign}
  &A^{\mathrm{cl}}_n =  \alpha_n\tilde{q}a_0^n,\quad B^{\mathrm{cl}}_n =  \beta_n\tilde{q}a_0^n,\\
  &A^{\mathrm{int}}_n =  \frac{1}{a^n}\big[\alpha_n f(a)^n+\frac{\beta_n}{f(a)^{n+1}}-\frac{1}{a^{n+1}}\big]\tilde{q}a_0^n,\\
  &B^{\mathrm{out}}_n =  b^{n+1}\big[\alpha_n f(b)^n+\frac{\beta_n}{f(b)^{n+1}}\big]\tilde{q}a_0^n,
\end{flalign}
\end{subequations}
where
\begin{widetext}
\begin{subequations}
\begin{flalign}
  &\alpha_n=\frac{\frac{\varepsilon_1}{\varepsilon_0}(n+1)(2n+1)(f(b)-b)f(a)^{n+1}}{a^n\big\{n(n+1)f(a)^{2n+1}(\frac{\varepsilon_1}{\varepsilon_0}a-f(a))(f(b)-b)+f(b)^{2n+1}[(n+1)b+nf(b)][\frac{\varepsilon_1}{\varepsilon_0}na+(n+1)f(a)]\big\}},\\
  &\beta_n=\frac{\frac{\varepsilon_1}{\varepsilon_0}(2n+1)f(b)^{2n+1}[(n+1)b+nf(b)]f(a)^{n+1}}{a^n\big\{n(n+1)f(a)^{2n+1}(\frac{\varepsilon_1}{\varepsilon_0}a-f(a))(f(b)-b)+f(b)^{2n+1}[(n+1)b+nf(b)][\frac{\varepsilon_1}{\varepsilon_0}na+(n+1)f(a)]\big\}}.
\end{flalign}
\end{subequations}
\end{widetext}
For the ideal situation $f(a)=0$, $f(b)=b$, the results are simplified to $\alpha_n=0$ and $\beta_0=\varepsilon_1/\varepsilon_0$, $\beta_n=0$ ($n\neq0$). Thus the potentials reduce to
\begin{subequations}\label{potentials ideal}
\begin{flalign}\label{potential1 ideal}
 &\psi^{\mathrm{out}}\ =\ \frac{q}{4\pi\varepsilon_0r},\quad \psi^{\mathrm{cl}}\ =\ \frac{q}{4\pi\varepsilon_0f(r)},
 \end{flalign}
 \begin{flalign}
 \begin{split}\label{potential2 ideal}
 \\\psi^{\mathrm{int}} = &  \frac{\tilde{q}}{\sqrt{r^2-2a_0r\cos\theta+a_0^2} }+\tilde{q}\left( \frac{\varepsilon_1}{\varepsilon_0}\frac{1}{f(a)}-\frac{1}{a} \right)\\
 &+\tilde{q}\sum_{n=1}^{\infty} \frac{n+1}{n}\frac{a_0^n}{a^{2n+1}} r^n P_n(\cos\theta),
 \end{split}
\end{flalign}
\end{subequations}
The field outside the cloak is precisely equal to the field produced by a point charge $q$ at the origin in vacuum. While the potential in the whole of the hidden area tends to infinite since the term $\varepsilon_1\tilde{q}/(\varepsilon_0f(a)) $ exists as $f(a)\rightarrow 0$.

We can further solve the Green's function to this set of boundary conditions in light of the above solutions. The Green's function $G(\vec{r},\vec{r}\,')$ is regarded as the potential produced by an unit point charge located at $(r',\, \theta',\, \phi')$. Thus $G(\vec{r},\vec{r}\,')$ can be directly transformed from  Eqs.~(\ref{potentials ideal}) through the substitution
\begin{subequations}
\begin{gather}
  q\ \rightarrow\ 1,\quad a_0\ \rightarrow\ r',\nonumber\\
  \cos\theta\ \rightarrow\ \cos\vartheta=\sin\theta\sin\theta'\cos(\phi-\phi')+\cos\theta\cos\theta'.\nonumber
\end{gather}
\end{subequations}
For an arbitrary distribution $\rho(\vec{r})$ of charges inside the cloak, the potential in the whole space is written as
\begin{equation}
  \psi(\vec{r})=\int G(\vec{r},\vec{r}\,')\rho(\vec{r}\,')\ dV',
\end{equation}
where the domain of integration is the whole hidden area. The potential outside the hidden area still takes the expression shown in Eq.~(\ref{potential1 ideal}), which means the fields outside is not affected by the distribution of charges inside the cloak but is only determined by the total charges $q$. Taking account of Gauss's theorem, it is actually not marvelous that the electrostatic cloak can not screen the outside space from the electric field of charged bodies inside, as a result the charged bodies can be detected outside the cloak. Nevertheless, apart from the total charges, no more information about the charge distribution can be detected, which is the least information allowed to gain under the restriction of Gauss's theorem. When the total charges tend to zero, no field can go out of the cloak which is in complete agreement with the result for an active source inside the cloak as its radiation frequency goes to zero \cite{Zhang2008PRL}.

The behavior of electrostatic shielding of the cloak is identical with a spherical conducting shell in the screened area no matter where the sources are located, inside or outside the cloak. However, the electric field in the source-existing area is different between the two systems. It is interesting to generalize the two systems as two different limitation of an unified system, a simple system with a dielectric sphere with permittivity $\varepsilon_1$ placed in another medium $\varepsilon_2$. For simplicity, we also consider that a point charge $q$ is located in the sphere with a distance $a_0$ from the center. The electric potential inside the sphere can be written as \cite{Batygin1978Academic}
\begin{equation}\label{potential of dielectric}
\begin{split}
 \psi^{\mathrm{int}} =\ &  \frac{\tilde{q}}{\sqrt{r^2-2a_0r\cos\theta+a_0^2} }+\tilde{q}\frac{\varepsilon_1-\varepsilon_2}{\varepsilon_1}\\
 &\cdot\sum_{n=0}^{\infty} \frac{n+1}{n+\frac{\varepsilon_2}{\varepsilon_0}(n+1)}\frac{a_0^n}{a^{2n+1}} r^n P_n(\cos\theta).
 \end{split}
\end{equation}

When $\varepsilon_2\rightarrow\infty$, the result is the same as the case of conducting shell. On the other hand, in the limit $\varepsilon_2\rightarrow0$, the result is identical with the solution for the cloak shown in Eq.~(\ref{potential2 ideal}) except the different constant term which in Eq.~(\ref{potential2 ideal}) is $\tilde{q}[ \varepsilon_1/(\varepsilon_0f(a))-1/a ]$ and which in the limit of Eq.~(\ref{potential of dielectric}) is the $n=0$ term $\tilde{q}\varepsilon_0/(\varepsilon_2a)$, while they both tend to infinity indeed. Actually, for spherical interface, only the radial component $\varepsilon_{\langle rr \rangle}$ has been used in the continuity of the normal component of $\vec{D}$. Thus Eq.~(\ref{potential of dielectric}) with $\varepsilon_2=0$ and the ideal cloak, whose radial component $\varepsilon_{\langle rr \rangle}$ tends to zero as $r\rightarrow a$, give the same result of field inside the hidden area.

For the system of ideal cloak, the electric field in the hidden area is finite, despite the potential in the whole hidden area is infinite. While the field in the cloak shell tends to infinity as $r\rightarrow a$. In addition, because the material of cloak is linear, the electrostatic energy of the system can be calculated by \cite{Jackson1999Wiley}
\begin{equation}
  W\ =\ \frac{1}{2}\int \psi^{\mathrm{int}}(\vec{r}\,')\rho(\vec{r}\,')\ dV'.
\end{equation}
According to the infinite constant inside the cloak, the electrostatic energy of the system tends to infinity for a nonzero total charges even if the term of self-energy is not included. The result reveals it would cost an infinite work to put charged bodies into the cloak, which is impossible in practice. However, if the total charges are zero, the infinite constant of potential tends to zero. The trouble of infinite work would disappear either.
\begin{figure}
\includegraphics[width=0.45\columnwidth,clip]{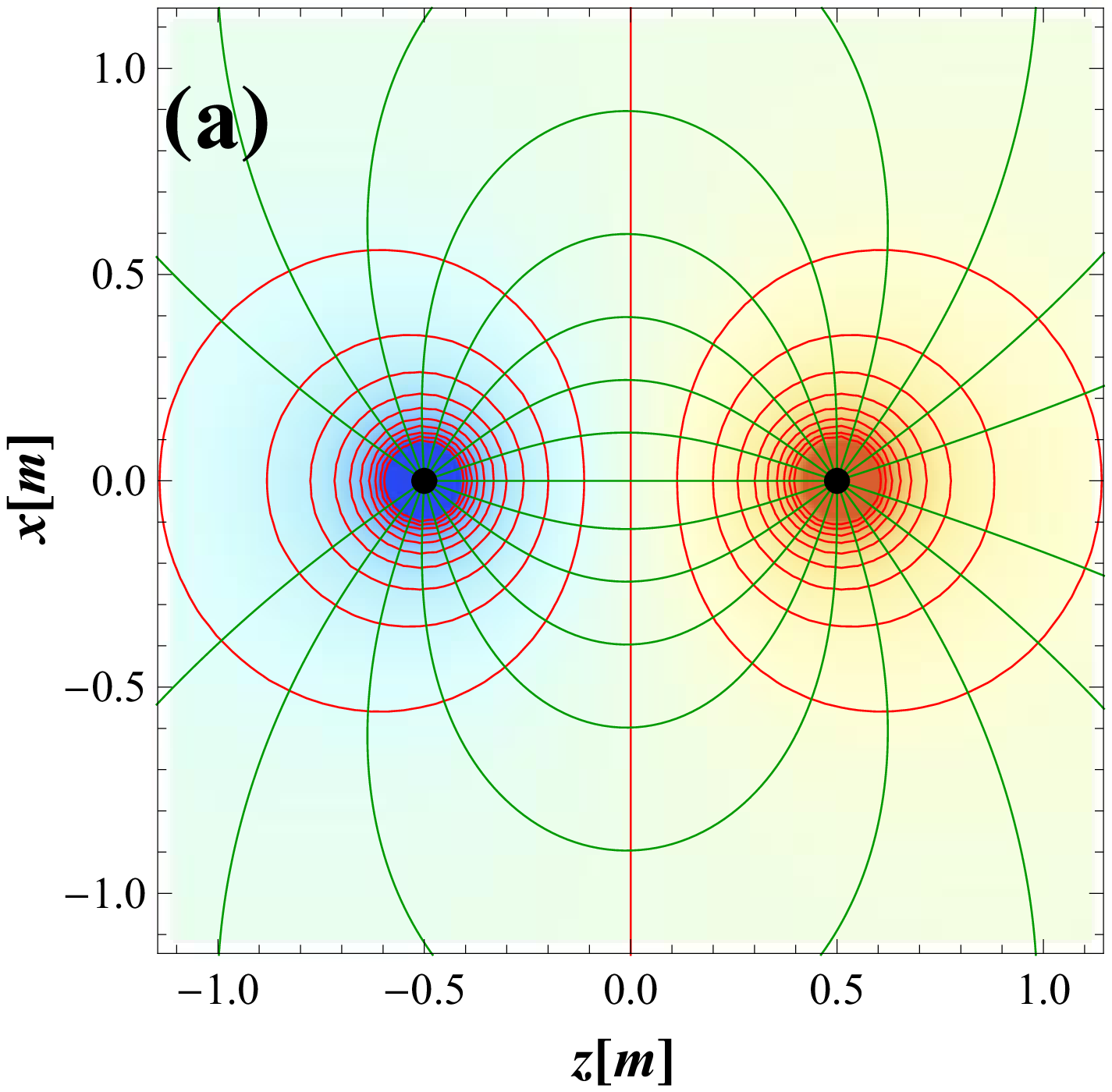}\label{fig2a}\\
\includegraphics[width=0.45\columnwidth,clip]{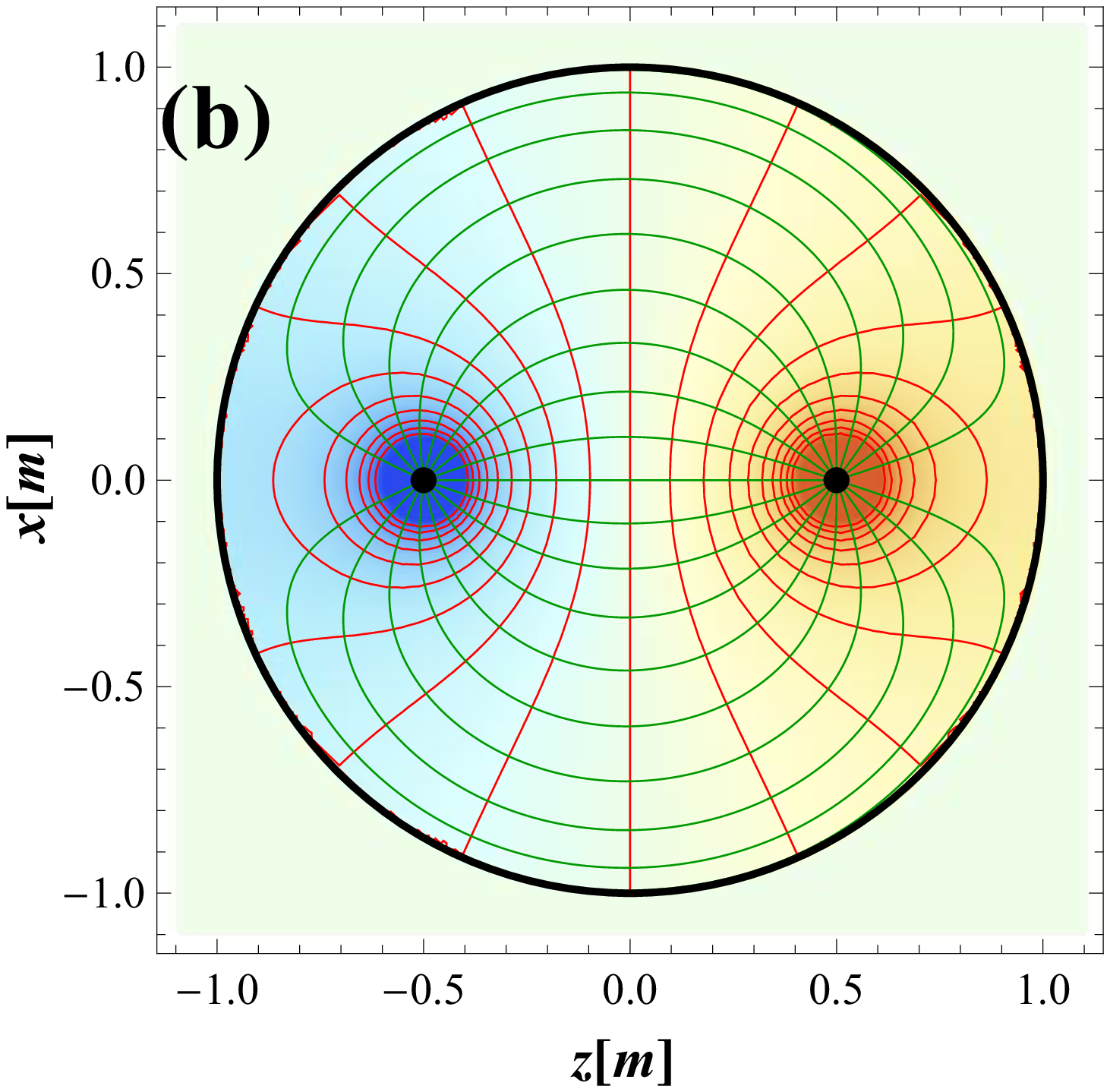}\label{fig2b}
\includegraphics[width=0.45\columnwidth,clip]{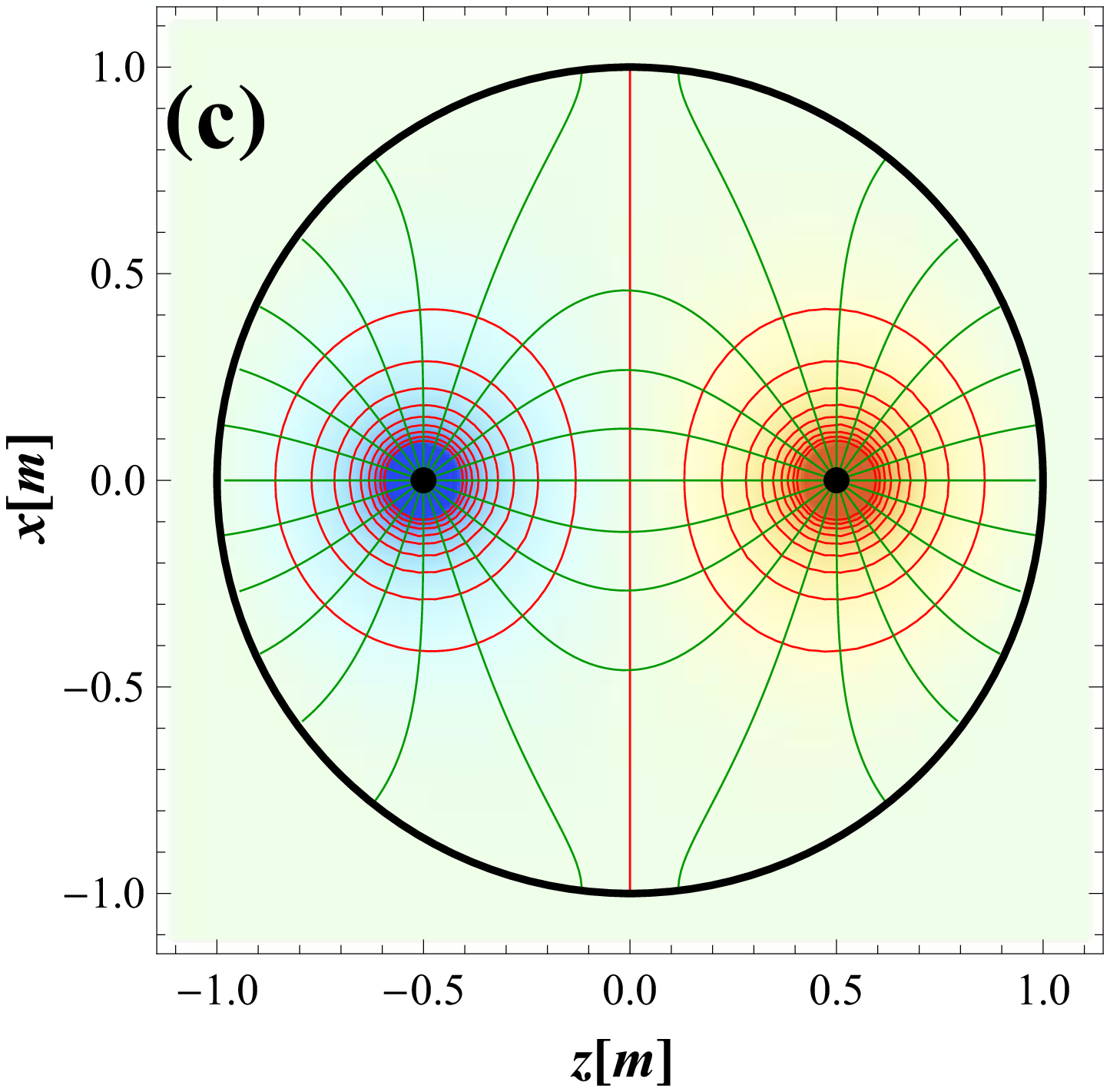}\label{fig2c}
\caption{\label{fig2}Electric field of two point charges with opposite quantity located at a distance $a/2$ from origin on positive and negative semiaxis of $z$ (a) in free space, (b) inside an ideal cloak, and (c) inside a spherical conducting shell.}
\end{figure}

Figure~\ref{fig2} shows an example of two point charges with opposite quantity located on positive and negative semiaxis of $z$ with a same distance $a/2$ from the origin respectively. When the two point charges are inside the ideal cloak, all electric lines of force do not intersect with the inner surface of the cloak, while the equipotential surface are orthogonal to the inner surface (Fig.~\ref{fig2}b). In contrast, as in the spherical conducting shell, all electric lines of force are orthogonal to the inner surface which itself becomes an equipotential surface (Fig.~\ref{fig2}c).

\section{surface voltage and equivalent surface magnetic current}

When checking the solution for an ideal cloak in Eqs.~(\ref{potentials ideal}), one can find that the potential is discontinuous across the inner interface of the cloak. For an arbitrary distribution of charges, the potential difference across the interface $r=a$ can be calculated by $\Delta\psi = \int\Delta G(a\hat{r},\vec{r}\,')\rho(r)\,'\,dV'$, where $\Delta G(a\hat{r},\vec{r}\,')$ is the difference of Green's function across the interface:
\begin{equation}
\begin{split}
 \Delta G(a\hat{r},\vec{r}\,') = & \frac{-1}{4\pi\varepsilon_1}\left\{ \frac{1}{\sqrt{a^2-2r'a\cos\vartheta+r'^2} }-\frac{1}{a}\right.\\
 &\left.+\sum_{n=1}^{\infty} \frac{n+1}{n}\frac{r'^n}{a^{n+1}} P_n(\cos\vartheta)\right\}.
\end{split}
\end{equation}
This discontinuity of potential comes from the terms $B^{\mathrm{cl}}_n f(r)^{-(n+1)}$ in $\psi^{\mathrm{cl}}$. For the ideal case $f(a)\rightarrow 0$, the coefficients  $B^{\mathrm{cl}}_n\rightarrow0$, so the terms of $f(r)^{-(n+1)}$  should be vanish in the cloak layer. While a meticulous calculation reveals that the limit $B^{\mathrm{cl}}_n f(a)^{-(n+1)}$ is towards to a finite quantity which acts as a surface voltage to balance the potential difference $\Delta\psi$ across the interface. The discontinuity also appears on the tangent component of electric fields. As we have pointed out in Fig.~\ref{fig2}, the electric fields are tangent to the interface $r=a$ on the inner side, yet are zero on the cloak side. Similarly, in the case of a cylindrical cloak with a transverse-electric (TE) incident wave, the tangent component of electric field is also discontinuous at the inner interface of the cloak \cite{Zhang2007PRB}. However, the discontinuities are caused by different reasons in the two cases. In cylindrical cloak, the tangent component of magnetic field $B_{\langle\phi\rangle}\rightarrow\infty$ at the inner surface $r=a$ of cloak, and the integral $\frac{d}{dt}\int_{a-0}^{a+0}B_{\langle\phi\rangle} dr  $ is equal to a finite value which acts as the surface magnetic displacement current to balance the difference of $E_{\langle z\rangle}$ in the Maxwell's Eq. $\oint \vec{E}\cdot d\vec{l}=-\frac{d}{dt}\int\vec{B}\cdot d\vec{s}$ for an infinitesimal contour across the interface \cite{Zhang2007PRB}. While, for the ideal case of the shielding phenomenon, the normal component of electric field $E_{\langle r\rangle}$ becomes a delta function compressed on the interface and contributes to the  surface voltage $\Delta\psi=-\int_{a-0}^{a+0}E_{\langle r\rangle} dr$ \cite{Zhang2008PRL}.

In the sense of dielectric materials, the surface voltage is caused by the infinite electric polarization of the material with $\varepsilon_{\langle rr \rangle}=0$ on the interface \cite{Zhang2008PRL}. However the surface voltage and the shielding effect of charges can also be caused by the the surface magnetic current in the dual superconductor. As we all known, the Meissner effect says that a superconductor expels magnetic fields from its interior and therefore the equivalent permeability of superconductor is equal to zero. Similarly, in dual superconductor model, magnetic monopoles are supposed to exist and the roles of electric and magnetic fields are interchanged. Moreover, the dual Meissner effect tries to expel electric fields out of the dual superconductors and gives a zero permittivity equivalently \cite{Ripka2005Springer}. In this case, the discontinuity of the tangent electric fields across the surface of the dual superconductor comes from the surface magnetic current $\vec{\alpha}$, which satisfies the boundary condition
\begin{equation}
  \vec{\alpha}\ =\ \left.\vec{E}^{\mathrm{int}}\times \hat{e}_r\right|_{r=a}.
\end{equation}
\begin{figure}
\includegraphics[width=0.45\columnwidth,clip]{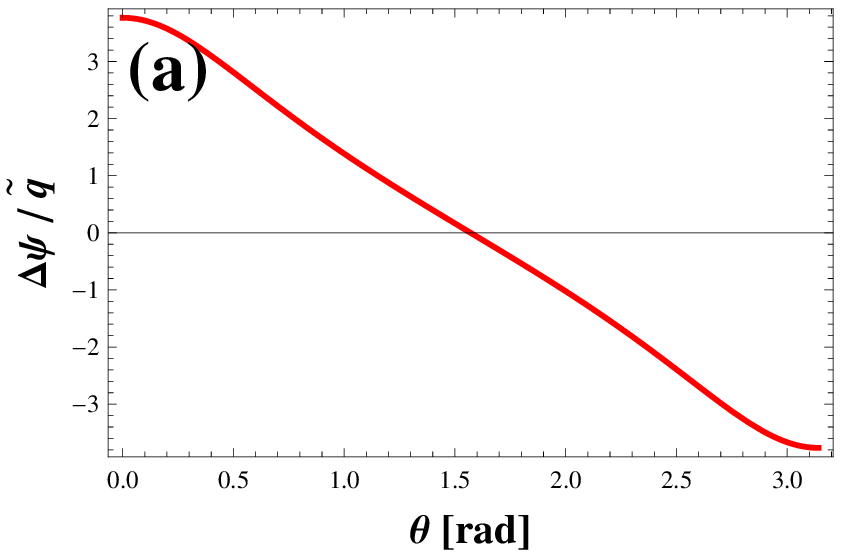}
\includegraphics[width=0.45\columnwidth,clip]{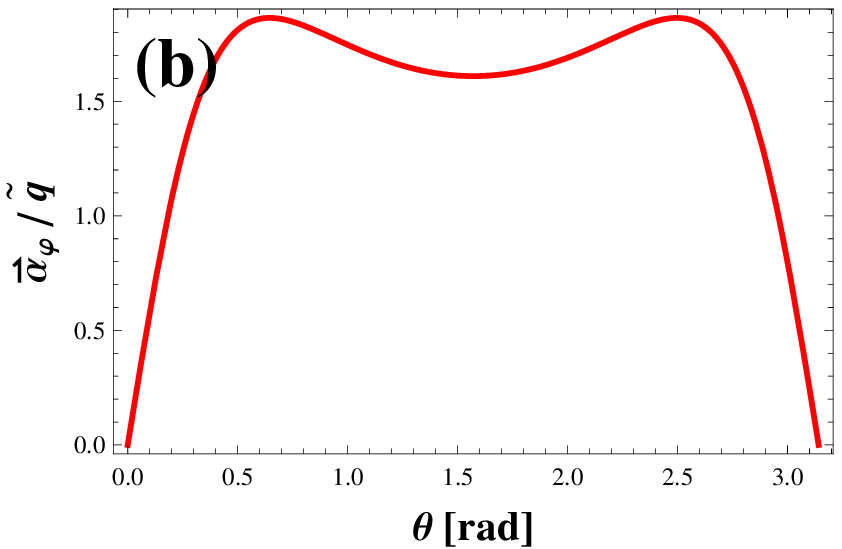}
\caption{\label{fig3}Distributions of (a) surface voltage $\Delta\psi$ and (b) surface magnetic current $\vec{\alpha}$ (only $\phi$ component exists) on the inner surface of the cloak for the charge distribution shown in Fig.~2 with normalized $\tilde{q}$.}
\end{figure}
\begin{figure}[b]
\includegraphics[width=0.45\columnwidth,clip]{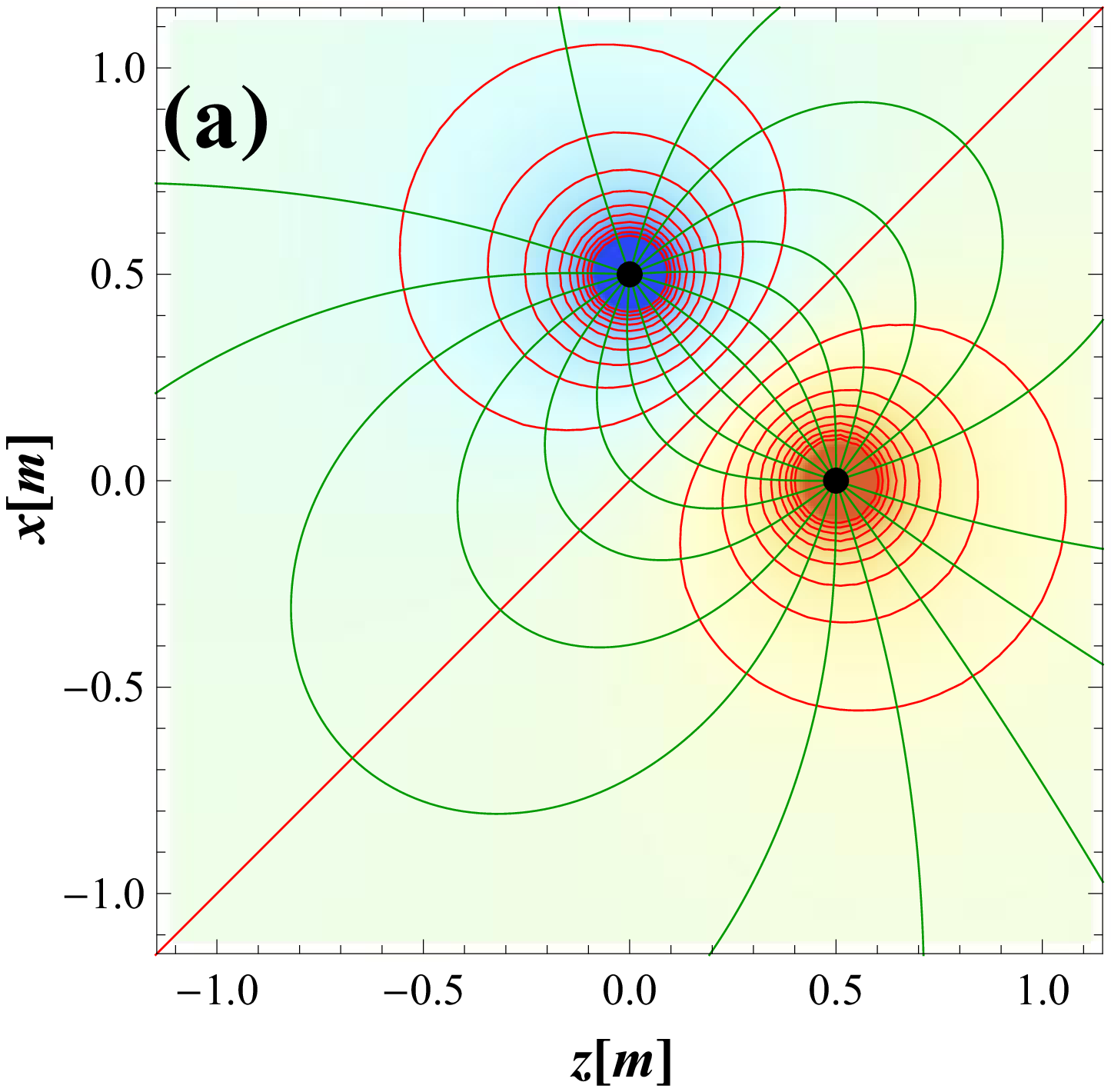}
\includegraphics[width=0.45\columnwidth,clip]{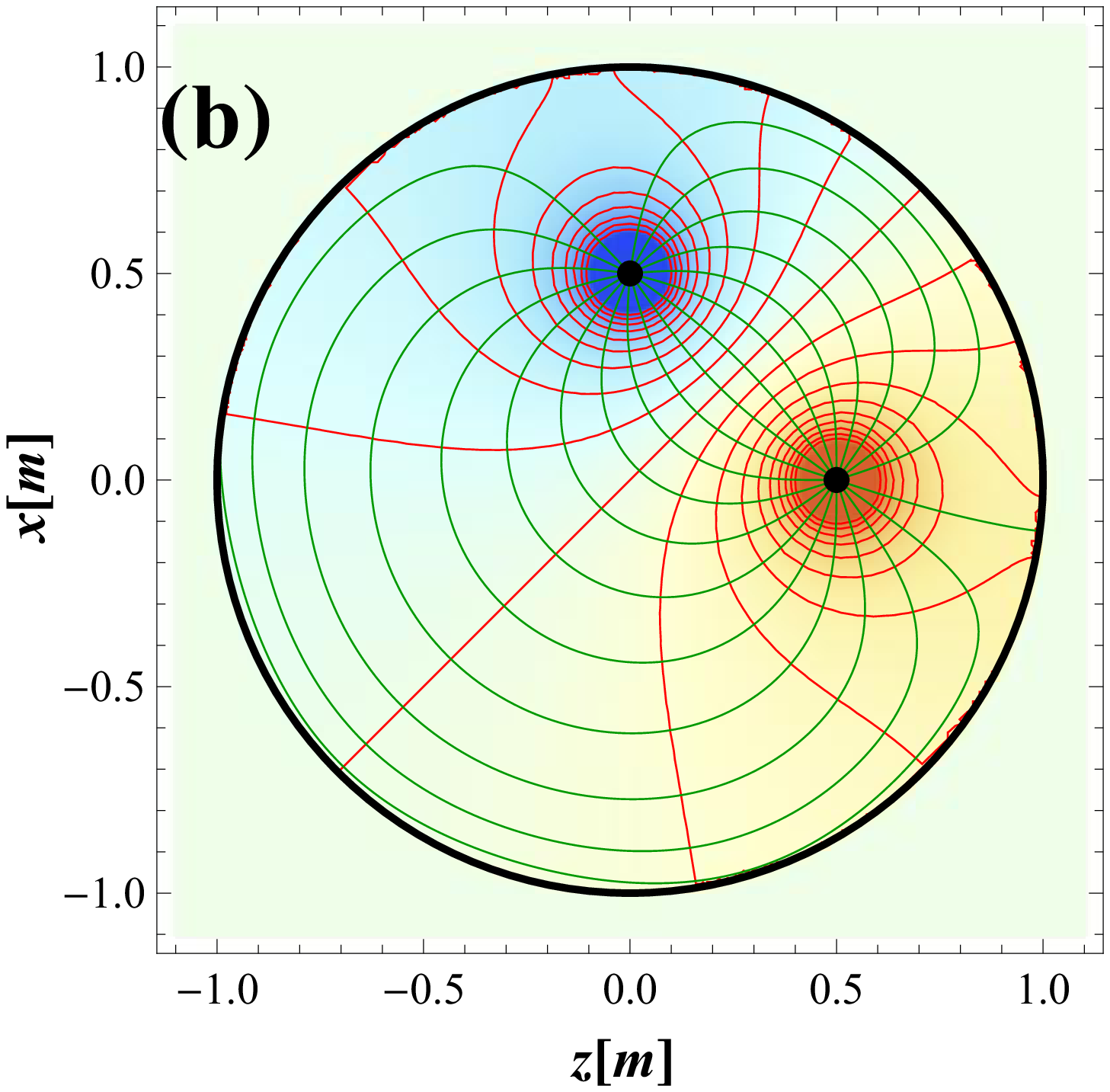}\\
\includegraphics[width=0.45\columnwidth,clip]{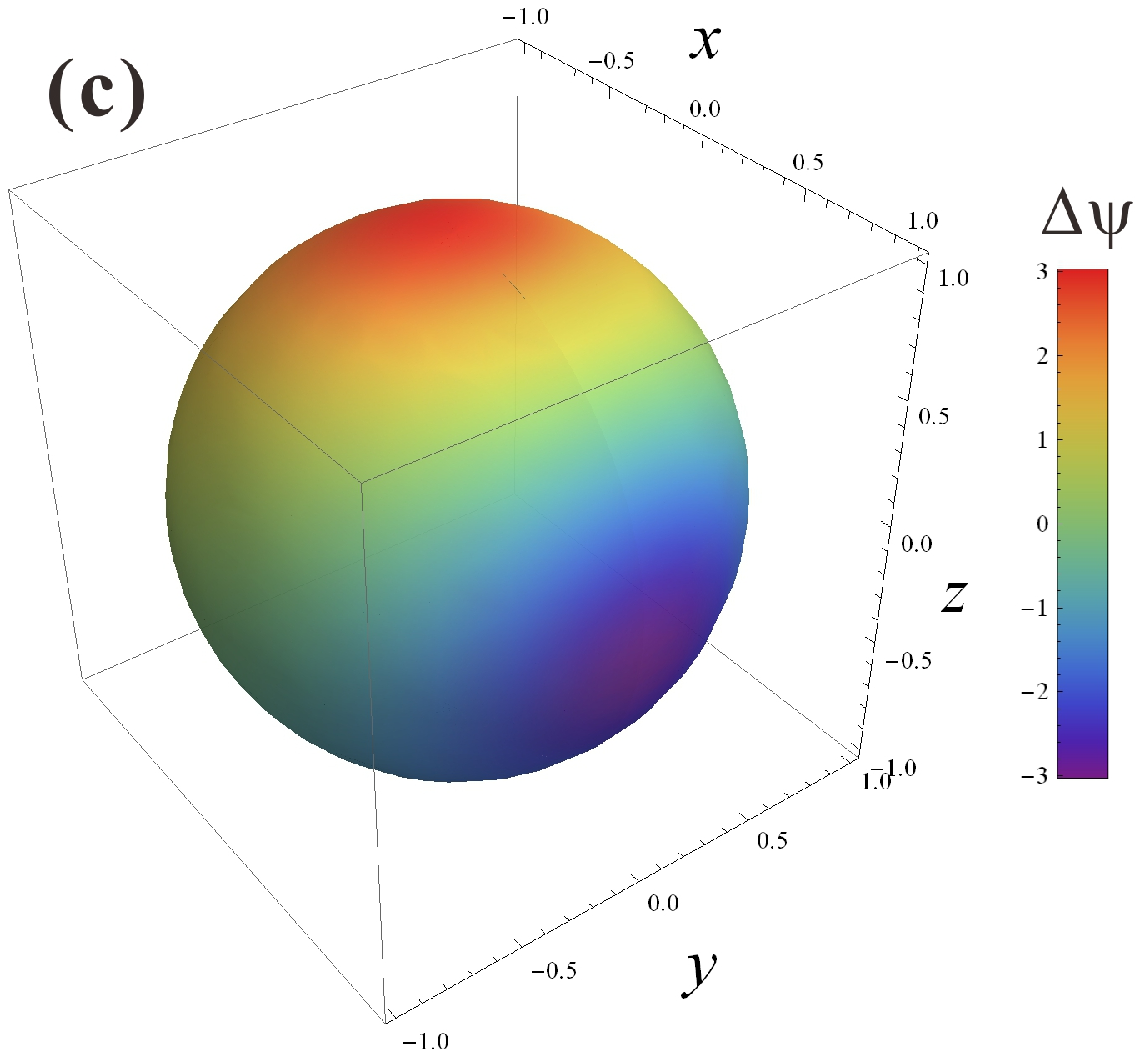}
\includegraphics[width=0.45\columnwidth,clip]{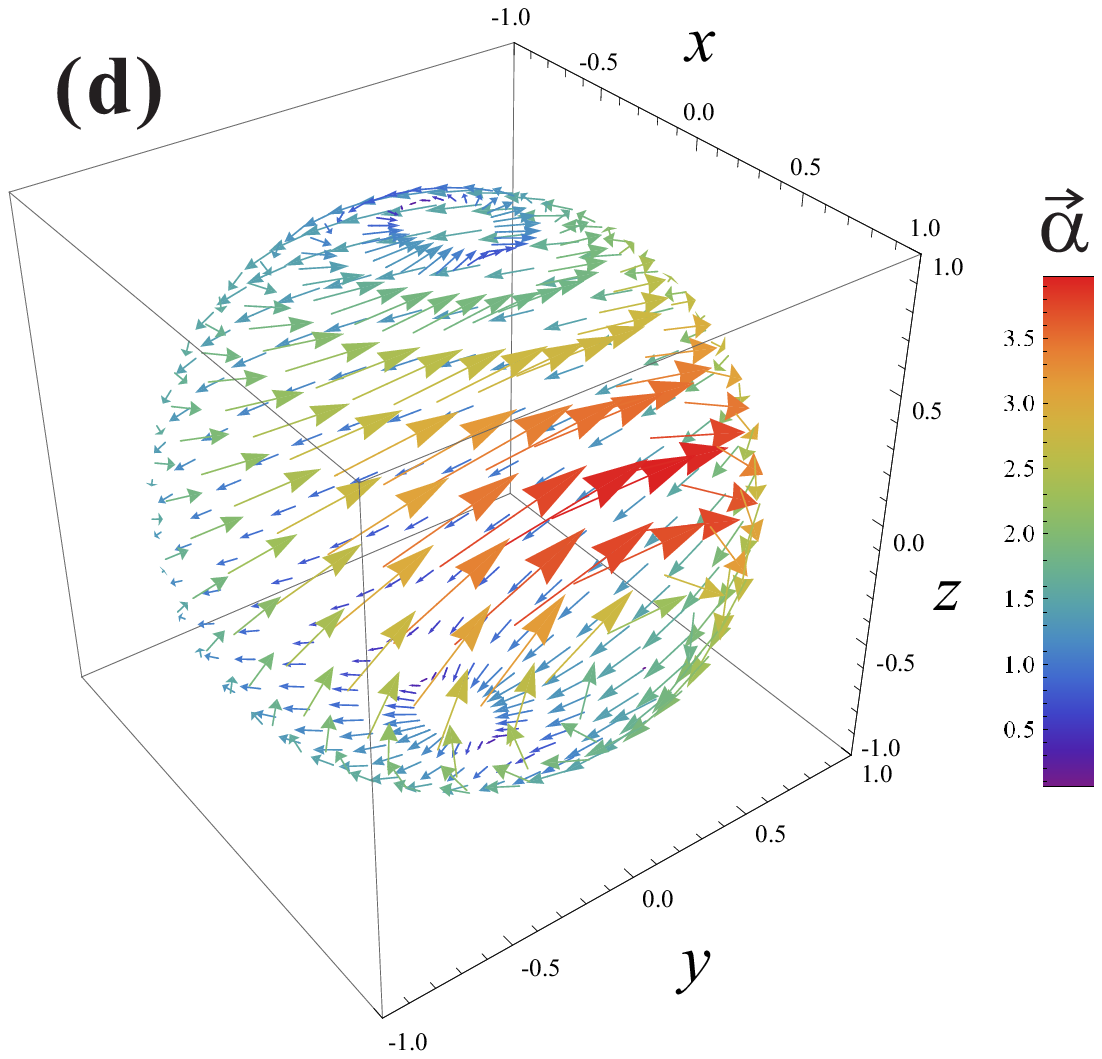}
\caption{\label{fig4}Electric field of two point charges with opposite quantity located at a distance $a/2$ from origin on $z$ axis and $x$ axis respectively (a) in free space, (b) inside an ideal cloak; and the distributions of (c) surface voltage $\Delta\psi$ and (d) surface magnetic current $\vec{\alpha}$ on the inner surface for the charge distribution with normalized $\tilde{q}$}
\end{figure}
As a result, the effect of electrostatic shielding of the ideal cloak is very like the charge confinement by the dual superconductors, so the inner surface of the cloak can be also interpreted as a dual superconductor layer. Although the medium of cloak is anisotropic, which is different from the simple dual superconductor, we still can construct the ideal electrostatic cloak with dual superconductive components as same as the way to construct magnetostatic cloak using superconductors suggested by B. Wood and J. Pendry \cite{Wood2007JPCM}. Figure~3 shows the curves of the surface voltage and the equivalent surface magnetic current varying with $\theta$ for the charge distribution shown in Fig.~2. Since the distribution is symmetric about $z$ axis, the voltage is also symmetric, and the surface current is along $\phi$ direction. Figure~4 presents another example of charge distribution (Fig.~4a,b), Fig.~4c shows the voltage distributing on the inner surface, and Fig.~4d shows the surface magnetic current on the inner surface.

\section{unideal cloak}

\begin{figure}
\includegraphics[width=0.45\columnwidth,clip]{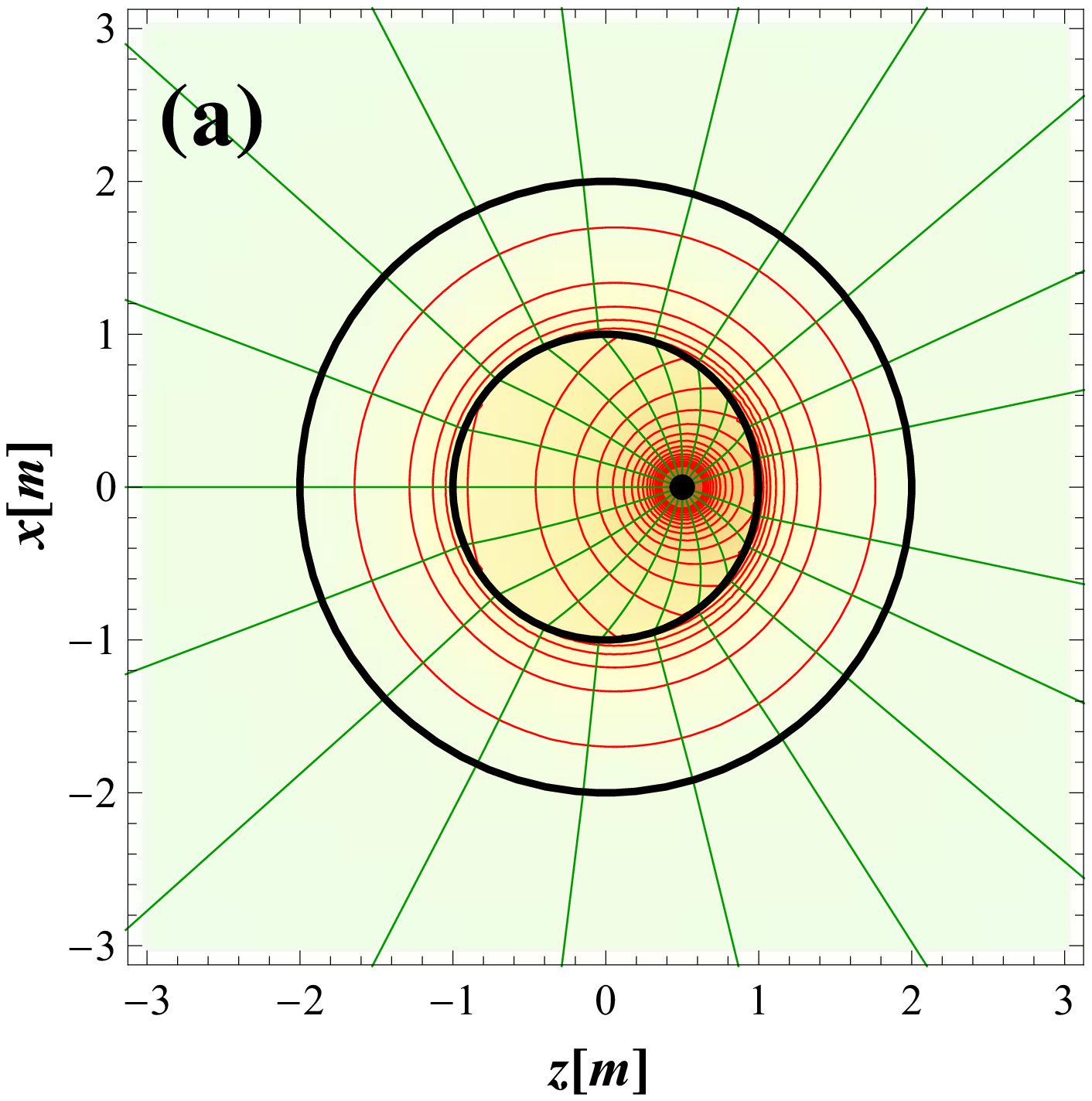}
\includegraphics[width=0.45\columnwidth,clip]{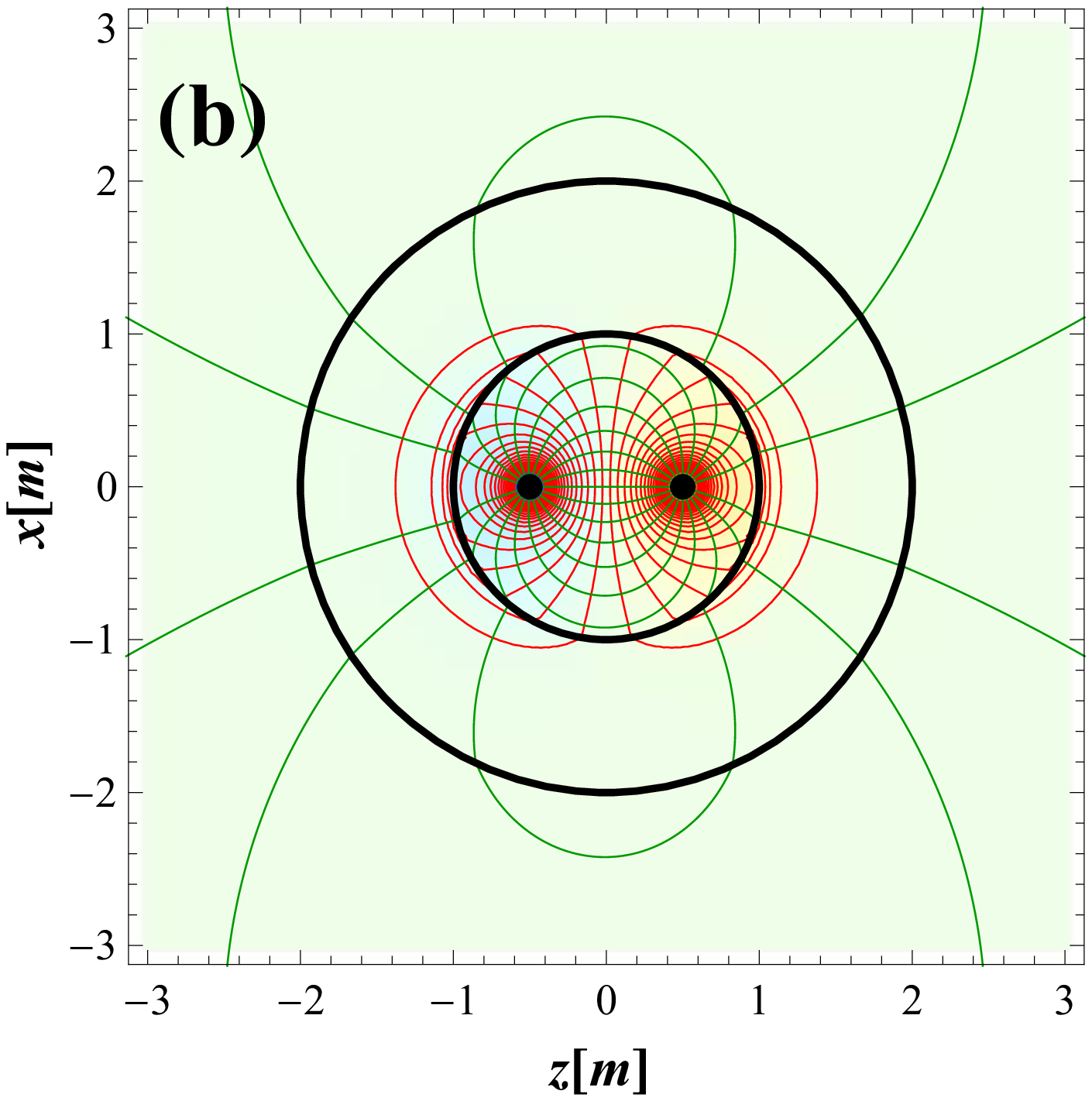}
\caption{\label{fig5}Electric field of (a) a point charge located on positive semiaxis of $z$ with a distance $a/2$ from origin, (b) two equal and opposite point charges located as same as in Fig.~\ref{fig2} for an unideal cloak constructed by the linear transformation function with $\delta=0.2$.  }
\end{figure}

In previous section, we have mentioned that the electric energy tends to infinite for the case of nonzero charges inside an ideal cloak, and it is impossible in practice. To ease this embarrassment, we would take into account the case of unideal cloak which is also obtained from the radial transformation $f(r)$. We still let $f(r)$ satisfy $f(b)=b$, however its zero point is not at $r=a$ but at $a'=(1-\delta)a$, when $\delta\rightarrow 0$ the cloak approaches to an ideal one. In this case, the potential inside the cloak no longer has the infinite constant term, and is continuous across the inner interface. A further calculation reveals that the unideal cloaks still have good property of electrostatic shielding both when the total charges is zero and nonzero. Figure~\ref{fig5} shows the case of unideal cloak designed by the linear transformation $f(r)=b(r-a')/(b-a')$ with $\delta=0.2$. For the case of a point charge located inside the cloak but not at the center, the fields outside the cloak still retain a highly spherical symmetry just as the fields generated by the point charge located at the center of the cloak, as shown in Fig.~\ref{fig5}a. For the case of two equal and opposite charges located inside the cloak, the fields are mainly concentrated inside the cloak, while the fields leaking out is extremely few, as shown in Fig.~\ref{fig5}b.

\begin{figure}
\includegraphics[width=0.9\columnwidth,clip]{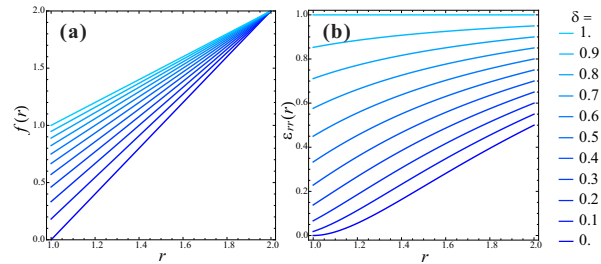}
\caption{\label{fig6}(a) Linear transformation $f(r)$ and (b) the corresponding radial component of permittivity $\varepsilon_{\langle rr\rangle}$ varying with $r$ in the cloak region under different value of $\delta$ from 0 to 1. }
\end{figure}

\begin{figure}[b]
\includegraphics[width=0.8\columnwidth,clip]{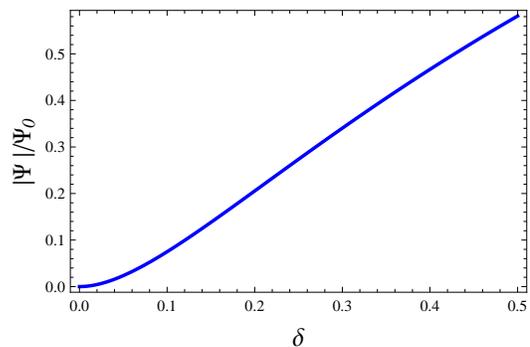}
\caption{\label{fig7}The absolute value of electric flux $|\Psi|$ divided by $\Psi_0$ flowing out of the hemispherical surface split by $z=0$ plane varying with $\delta$. }
\end{figure}

\begin{figure}
\includegraphics[width=0.9\columnwidth,clip]{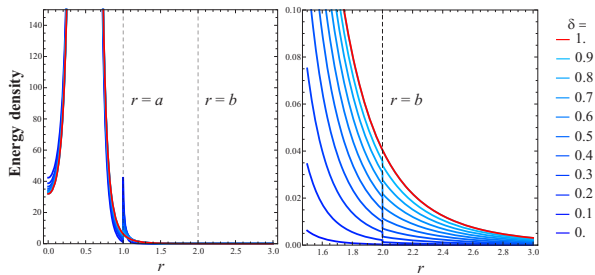}
\caption{\label{fig8}Electric field energy density varying on $z$ axis under different value of $\delta$, where the red line denotes the case of no cloak existing ($\delta=1$). }
\end{figure}

To measure the effect of shielding accurately, we consider the electric flux propagating out of the cloak. Concerning the system of two equal and opposite charges inside the cloak as shown in Fig.~\ref{fig2}b, Gauss' theorem tells the total flux out of the cloak is zero. However, if we calculate the flux out of two hemispherical surfaces of a sphere $r=r_0>b$ split by $z=0$ plane respectively, the pair of flux must have an equal magnitude $|\Psi|$ but opposite sign because of the symmetry of charge distribution. Thus the absolute value of electric flux out of each hemisphere would be a suitable quantity to measure the effect of shielding for this system. We still consider the linear radial transformation $f(r)$ with different $\delta$ (Fig.~\ref{fig6}a) and corresponding permittivities (Fig.~\ref{fig6}b). In this situation, $|\Psi|/\Psi_0$ varying with $\delta$ is shown in Fig.~7, where $|\Psi|$ is the absolute value of flux  flowing out of the hemisphere and $\Psi_0$ is the absolute value of flux out of the hemisphere with no cloak existing. The slope of $|\Psi|/\Psi_0$ varying with $\delta$ is equal to zero as $\delta\rightarrow 0$. The result manifests that a nearly ideal cloak would still realize a good effect of shielding. In fact, the derivative of $\varepsilon_{\langle rr\rangle}$ of ideal cloak with respect to $r$ is equal to zero for arbitrary $f(r)$, therefore a nearly ideal cloak with small value of $\delta$ would always have a good behavior of shielding. The effect of shielding can be measured from another aspect of energy density $\frac{1}{2}\vec{E}\cdot\vec{D}$, as shown in Fig.~\ref{fig8}. We can see that the energy density outside a nearly ideal cloak tends to zero when $\delta$ is very small and is much smaller than the case of no cloak existing. Along with the increase of $\delta$, the energy density increases nearly stable and goes to the limit of no cloak existing when $\delta\rightarrow 1$. To sum up, the slight breaking of the perfection would not change the property of electrostatic shielding for the cloak significantly.

\section{conclusion}
To summarize, we have verified that electrostatic spherical cloak takes the same form of permittivity as the invisibility cloak under nonzero frequency, and demonstrated its behavior of electrostatic shielding is identical with a spherical conducting shell in the screened region. If the electrostatic sources are outside the cloak, the field can not propagate into the screened region inside the cloak, on the other hand, if the charges with arbitrary distribution are inside the cloak, the field outside the cloak is just as the field generated by a point charge located at the center of the cloak, so the only information which can be detected is the total charges inside the cloak. For ideal case, the potential across the inner interface of the cloak is not continuous in that $\varepsilon_{\langle rr\rangle}=0$ causes the infinite polarization on the inner interface. However, $\varepsilon=0$ can be also interpreted as the property of a dual superconductor, and the behavior of shielding are also very like the property of charge confinement caused by dual Meissner effect. Another problem existing in ideal case is the infinite field energy when the total charges inside the cloak is not zero. Nevertheless, the problem no longer exists for unideal case, in addition the nearly ideal cloak also have a good effect of electrostatic shielding.


\begin{acknowledgments}
  This work was supported in part by NSF of China (Grants No. 11075077).
\end{acknowledgments}

\bibliography{ESIC}






\end{document}